%% file: paper.tex
\pdfoutput=1 

\documentclass[journal]{IEEEtran}
\ifCLASSINFOpdf
\else
\fi
\usepackage{epsfig,endnotes}

\newcommand{\TITLE}{Reducing Storage in Large-Scale Photo Sharing Services using Recompression}

\newcommand{\PAGENUMBERS}{yes}       

\newcommand{\showComments}{yes}
\newcommand{\comment}[1]{}

\usepackage{ifthen}
\ifthenelse{\equal{\PAGENUMBERS}{yes}}{%
\usepackage[nohead,
            left=0.75in,right=0.75in,top=1.0in,
            footskip=0.5in,bottom=1in,     
            columnsep=0.25in
            ]{geometry}
}{%
\usepackage[noheadfoot,left=0.75in,right=0/75in,top=1.0in,
            footskip=0.5in,bottom=1in,
            columnsep=0.25in
	    ]{geometry}
}

\newcommand{\paraspace}{\vspace{0.05in}}
\newcommand{\parab}[1]{\paraspace\noindent{\bf #1} }
\newcommand{\parae}[1]{\paraspace\noindent{\em #1} }

\usepackage{subcaption}
\usepackage{booktabs}
\usepackage{color}
\usepackage{colortbl}
\usepackage{float}                           
\usepackage{graphicx}
\usepackage{soul}
\usepackage{fancyvrb}
\usepackage{multirow}
\usepackage{epsfig}
\usepackage{xspace}
\usepackage{textcomp}  
\usepackage{amsmath}
\usepackage{soul}

\begin{document}
%
\title{\TITLE}
%
%
%

\author{Xing Xu, Zahaib Akhtar, Wyatt Lloyd, Antonio Ortega, Ramesh Govindan}

\maketitle

\input{commands}

\begin{abstract}
\input{abstract}
\end{abstract}

\begin{IEEEkeywords}
JPEG, image compression, photo-sharing service, photo storage.
\end{IEEEkeywords}

\input{intro}
\input{design}
\input{recompression}
\input{eval}
\input{related}
\input{concl}
\IEEEpeerreviewmaketitle


\ifCLASSOPTIONcaptionsoff
  \newpage
\fi



%
\bibliographystyle{IEEEtran}
\bibliography{ref}

%








\end{document}

%% file: commands.tex
%
%

\newcommand{\sysname}{\textsc{ROMP}\xspace}
\newcommand{\lossyname}{\textsc{L-ROMP}\xspace}
\newcommand{\aggrename}{\textsc{L-ROMP-Aggressive}\xspace}
\newcommand{\names}{\textsc{ROMPs}\xspace}

\newcommand{\tld}{\texttildelow}
\newcommand{\realnewpage}{\onecolumn\twocolumn}
\newcommand{\warnpoint}{\noindent\\\noindent{\color{red}WARNING - Draft still under heavy revision past this point.\label{warnpoint}}}

\newcommand{\note}[2]{
    \ifthenelse{\equal{\showComments}{no}}{{\color{#1}[#2]}\\}{}
}

\definecolor{eblue}{RGB}{0, 0, 139}
\definecolor{mblue}{RGB}{0, 147, 175}
\definecolor{rgreen}{RGB}{0, 112, 60}
\definecolor{worange}{RGB}{245, 128, 37}
\definecolor{xred}{RGB}{200, 50, 50}
\newcommand{\ao}[1]{\note{eblue}{AO: #1}}
\newcommand{\aod}[1]{}
\newcommand{\za}[1]{\note{mblue}{ZA: #1}}
\newcommand{\zad}[1]{}
\newcommand{\rg}[1]{\note{rgreen}{R: #1}}
\newcommand{\rgd}[1]{}
\newcommand{\wl}[1]{\note{worange}{W: #1}}
\newcommand{\wld}[1]{}
\newcommand{\xx}[1]{\note{xred}{XX: #1}}
\newcommand{\xxd}[1]{}

\newcommand{\syma}{\emph{runsize}\xspace}
\newcommand{\symas}{\emph{runsize}s\xspace}
\newcommand{\symb}{\emph{amplitude}\xspace}
\newcommand{\symbs}{\emph{amplitude}s\xspace}
\definecolor{fillorange}{RGB}{180, 90, 30}
\newcommand{\fillin}[1]{{\color{fillorange}{#1}}}

\newcommand{\twocolumnfigurewidth}{.85\linewidth}
\newcommand{\subfigurewidth}{.46\textwidth}

\newcommand{\shortenitemize}{\vspace{-0.2cm}}
\newcommand{\shortenfigurecaption}{\vspace{-0.4cm}}
\newcommand{\shortentablecaptiona}{\vspace{-0.3cm}}
\newcommand{\shortentablecaptionb}{\vspace{-0.6cm}}

%% file: abstract.tex
The popularity of photo sharing services has increased dramatically in
recent years.  Increases in users, quantity of photos, and
quality/resolution of photos combined with the user expectation that photos
are reliably stored indefinitely creates a growing burden on the
storage backend of these services. We identify a new opportunity for
storage savings with application-specific compression for
photo sharing services: \emph{photo recompression}.
  
We explore new photo storage management techniques that are fast so
they do not adversely affect photo download latency, are
complementary to existing distributed erasure coding techniques, can
efficiently be converted to the standard JPEG user devices expect, and
significantly increase compression. We implement our photo
recompression techniques in two novel codecs, \sysname and \lossyname.
\sysname is a lossless JPEG recompression codec that compresses
typical photos 15\% over standard JPEG. \lossyname is a lossy JPEG
recompression codec that distorts photos in a perceptually
un-noticeable way and typically achieves 28\% compression over
standard JPEG. We estimate the benefits of our approach on Facebook's photo stack and find that our approaches can reduce the photo storage by 0.3-0.9$\times$ the logical size of the stored
photos, and
offer additional, collateral benefits to the photo caching stack,
including 5-11\% fewer requests to the backend storage, 15-31\% reduction
in wide-area bandwidth, and 16\% reduction in external bandwidth.

%% file: intro.tex
\section{Introduction}
\label{sec:intro}


\IEEEPARstart{I}{n} recent years, there has been a dramatic growth in the popularity of
large-scale photo sharing services such as Facebook, Flickr, and
Instagram. As of 2013, Facebook alone had 350 million photo uploads
per day~\cite{facebook_whitepaper}.  The storage footprint of these
services is already significant; Facebook stored over 250 billion
photos according to the same 2013 report~\cite{facebook_whitepaper}.
Furthermore, the commoditization of high-quality digital cameras in
mobile devices has created three trends that each increases the
footprint of these services: people are taking more photos, at higher
resolution, and at higher quality.  These trends combined with the
user expectation that photos are stored indefinitely results in an
ever growing storage footprint.

Today, photos uploaded to photo sharing services predominantly use
the JPEG~\cite{jpeg_still} standard, which already compresses images
by leveraging properties empirically observed in natural images. Despite this compression, as photo
sharing service scale, additional tools for managing photo
storage become necessary. 

One technique, already used in Facebook's f4 system~\cite{f4}, is
distributed erasure coding. This reduces the storage footprint of a
service by replacing the redundant copies of data that were used for
fault tolerance and load balancing requests with combined parity
information for multiple sets of data. Another prominent technique
explored by the storage systems community is deduplication
(Section~\ref{sec:related}). To our knowledge, this has not been
applied to images, in part because it is not clear if, after JPEG's
compression, duplicate elimination
is likely to provide significant benefit. For a similar reason, generic object
compression techniques---e.g., gzip, bzip2---are unlikely to produce
additional savings in storage.

Two other tools are available to large-scale photo sharing services:
image resizing and reducing JPEG's quality parameter (this is a lossy
transformation that re-quantizes information to enable better
compression). As cameras and displays move towards higher resolution,
users will likely want to view larger images, so the benefit of 
image resizing is limited. Moreover, as we show later, reducing JPEG's
quality parameter by re-quantization can introduce significant error
(uploaded images have already been quantized once when the JPEG was
generated at the source).

In this paper, we focus on the problem of \emph{image recompression}:
taking uploaded compressed images, and recompressing them by
taking advantage of the special characteristics of 
large-scale
photo sharing systems.  Recompressing images reduces the logical size
of the stored corpus and thus is complimentary to more generic
techniques like distributed erasure coding.  

There are three primary challenges for recompression schemes. The
first is finding opportunities for additional compression given that
images are already compressed. The second is to introduce minimal
error if lossy recompression is used, a property that reducing JPEG's quality
parameter does not have.  The third stems from ensuring compatibility
with client devices.  Maintaining compatibility requires clients
receive images in the JPEG format their devices understand and this in
turn requires decompression from the storage format back into JPEG on
the download path.  This means that decompression should be fast (or,
equivalently, have \emph{low complexity})---i.e., it should take
$<$0.1s and ideally $<$50ms---because it adds to the user-perceived
download latency of a photo sharing service (Section~\ref{sec:integrating}).

\parab{Contributions.}
Our key insight that enables this recompression is that large-scale photo sharing
services represent a different domain than those for which image
formats were designed. The \emph{large-scale} aspect enables further
compression of already compressed photos. At a high level, our approach
decouples the size of the codec tables used for encoding and
decoding many types of images. This decoupling
enables much richer and larger codec tables than that are practical for
individual files, \emph{i.e.}, the size of these large tables can be
amortized over the large-scale storage and thus become negligible. In addition, in contrast to the traditional compression
setting that considers individual images sent between a distinct
sender and receiver, recompression for photo sharing services involves
many photos that are compressed and decompressed by the same entity.
These insights leads to our first recompression scheme, Recompression
Of Many Photos (\sysname, Section~\ref{sec:recompression}).

\sysname achieves high recompression rates by replacing the small
coding tables that are stored with each image (or used by encoders/decoders as default) in traditional schemes
with a single large coding table that is not stored with the images.
The co-location of compression and decompression allows \sysname to
avoid storing the coding table with each image.  Instead, it stores
the table in the memory of machines on the download path.  This in
turn allows \sysname to use a much larger coding table than would be practical 
for individual images. Such significant increase in coding table sizes can be amortized over the billions of photos for large-scale storage. \sysname achieves low complexity by keeping the coding
table in memory on the download path of a photo sharing service and
only making a single pass over an image.

\sysname provides high compression and low complexity for
recompressing many photos and is lossless, i.e., recompressing a JPEG
into \sysname and back produces a bit-wise identical image.  For further compression
gains, one could think of using other existing image compression algorithms
(e.g., JPEG 2000 \cite{jpeg2000}). 
Note, however, that JPEG is the {\em de facto} standard for ``raw'' image
representation, that is, most images are first stored as high quality JPEG 
files on a digital camera. 
Then, applying other compression schemes would require decoding JPEG images (at
upload) and  generating again JPEG images (at download), as many potential clients only support JPEG, which would incur a significant complexity penalty. As a
result, 
the standard method for recompressing images is decreasing
JPEG's quality factor, but this increases distortion due to double-rounding of image
coefficients. Our Lossy-Recompression Of Many Photos (\lossyname) is
designed to reduce bit-rate in JPEG images and avoids this double-rounding problem. \lossyname
amplifies the compression gain of \sysname, adds no addition complexity to
decompression, and is perceptually nearly lossless. \sysname and \lossyname have been published in \cite{icassp}. This paper includes more results for evaluation of \sysname and \lossyname.\footnote{\cite{icassp} presents the compression ratio results of Figures \ref{fig:fivek_tradeoff-1} and \ref{fig:fivek_tradeoff} (but not complexity results), and complexity results of Figure \ref{fig:complexity_quality} for quality parameter 75 only. These are the only overlap in results presented in the two papers.} More importantly, this paper focuses more on leveraging \sysname and \lossyname in photo sharing services, but not the codec itself, and presents system level evaluation.

In addition to a storage backend, photo sharing services typically
deploy a photo-caching stack~\cite{huang13fbcdn}.  The stack is a set
of caching layers that are progressively smaller and closer to
clients.  The goals of the stack are three-fold: to reduce the load on
the storage backend, to reduce the amount of external bandwidth needed
to deliver photos, and to decrease user-perceived download latency.
The most straightforward deployment of \sysname would place the
decompression step between the storage backend and the caching stack.
We find, however, that doing the decompression inside the caching
stack would improve each of the three goals.  We term these ``collateral
benefits'' because they are in addition to the reduction in storage cost.

Our evaluation explores the compression and complexity of \sysname and
\lossyname across a variety of image qualities and resolutions.  We find that
\sysname and \lossyname robustly provide high
compression---approximately 15\% and 28\% over JPEG Standard
respectively---and low complexity---less than $60$ms decompress time. This translates to 13\% and 26\% compression
over JPEG Optimized\footnote{JPEG Optimized is a lossless compression technique for JPEG Standard format, which provides additional compression with negligible overhead. It is a more popular JPEG format.}. These storage
savings multiply when integrated into the photo-sharing
services. Because photos are replicated or erasure coded in the backend
for fault-tolerance, \sysname and \lossyname can reduce the storage footprint by
$0.5\times$--$0.9\times$ the logical size of
stored photos using the lossless/lossy codecs respectively. We also estimate the collateral benefits of deploying our schemes in
Facebook's photo caching stack. These would, for example, include 5\%--11\% fewer
requests to the backend storage, a 15\%--31\%
reduction in wide-area network bandwidth, and a 500ms
decrease in 99$^{\textrm{th}}$ percentile  photo download
latency.

%% file: design.tex
\section{Integrating Recompression into Photo Sharing Services}
\label{sec:integrating}

In this section, we discuss the benefits, for large-scale photo
sharing services, of recompressing uploaded images and how to
integrate the functionality provided by \sysname or \lossyname into
these photo sharing services. Our discussion is especially informed by
the descriptions of Facebook's photo sharing stack~\cite{haystack, f4,
  huang13fbcdn}, and our measurements of it.

From a systems point of view, \sysname and \lossyname each consist of
two conceptual modules: a compression module that recompresses an
uploaded image, and a decompression module that decompresses images
before download.

\begin{figure}[t]
\centering
\noindent\includegraphics[width=.9\columnwidth]{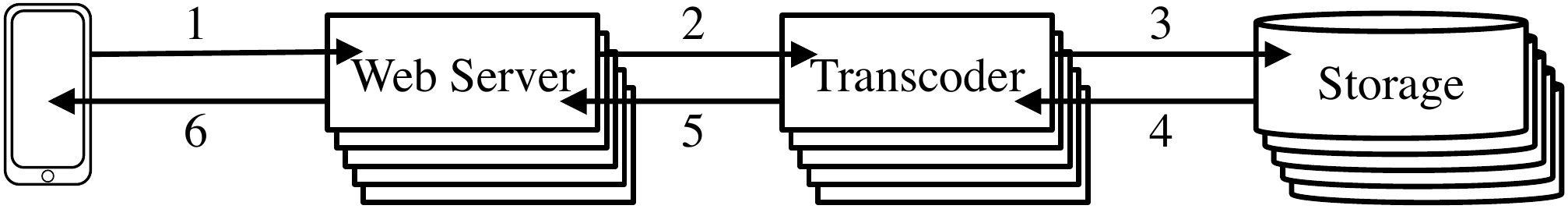}
\caption{A typical upload path (1--6) for a large-scale photo sharing
  service.  Photos are sent synchronously from users to backend
  storage via front-end Web servers and a transcoding tier.}
\label{fig:upload_path}
\end{figure}

\parab{Compressing on Upload.}
A typical upload path for a large-scale photo sharing service is shown
in Figure~\ref{fig:upload_path}.  Users upload photos to the service
synchronously, i.e., the user waits until the photo is safely stored
in the backend storage.  The user's device initially sends the
photo to a front-end Web server that handles incoming requests from
clients (1).  That Web server then sends the photo to a transcoder
machine in the transcoding tier (2).  The transcoder then sends the
photo to the backend storage tier (3), and, once the photo is safely
stored, acknowledgements flow in the reverse direction (4--6).

The transcoder sits on the upload path and canonicalizes photos before
storing them.  This canonicalization typically involves resizing
photos and/or reducing the JPEG image quality. The JPEG standard permits the reduction
of image quality using a scalar \emph{quality parameter}.  This
quality reduction reduces the storage required for images but can
introduce undesirable perceptual artifacts. Resizing photos gives them
standard sizes and ensures that photo storage growth is consistent and
predictable.  For instance, this prevents the release of a new popular
phone model with a higher resolution camera from increasing the
required storage per photo.  Reducing image quality to a fixed factor
also keeps storage growth consistent and predictable for similar
reasons. Both of these
transformations are \emph{lossy}, a topic we cover and explore in more
depth in Section \ref{sec:romp-lromp}.

Given that the transcoding tier is already doing image processing on
uploaded photos it is a natural place to also do recompression.
Doing recompression here requires that compression must have low
\emph{complexity} (time taken to recompress the images). In
particular, the complexity of recompression must be comparable to,
or less than, the complexity of resizing or reducing image quality. 

An alternative place to integrate recompression of photos would be off
the upload path.  This would require storing the photos in their
initial format and then recompressing them during off-peak periods.
This might be feasible but adds complexity to the entire path: these
recompression operations may need to be carefully scheduled, and the
download path (described below) needs to be able to retrieve images
before they have been scheduled for recompression (since users expect
images to be available immediately). For these reasons, we do not
consider this alternative in this paper.

\begin{figure}[t]
\centering
\noindent\includegraphics[width=.9\columnwidth]{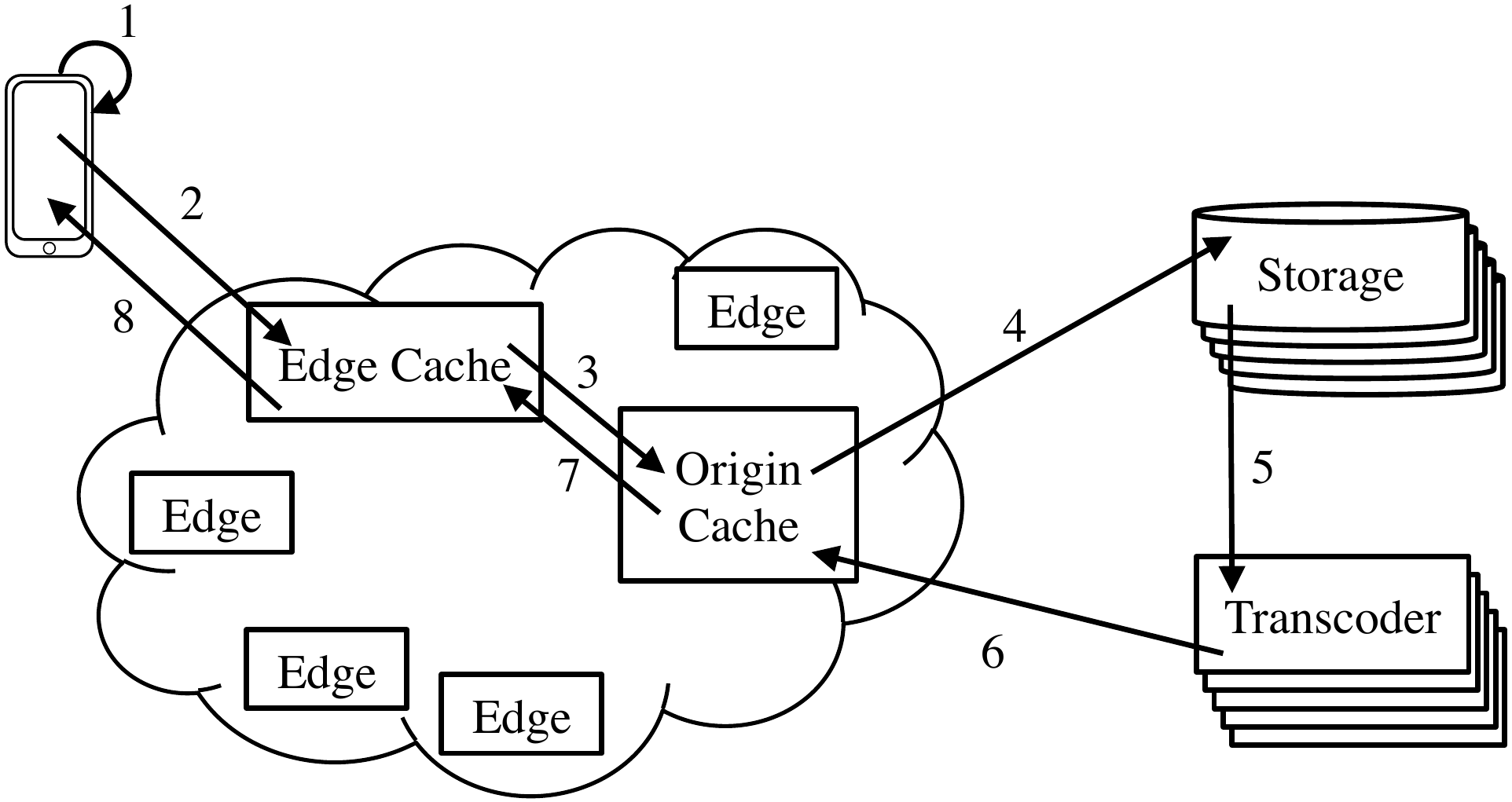}
\caption{A typical download path (1, 1-2-8, 1-2-3-7-8, or 1--8) for a
  large-scale photo sharing service.  A photo is returned from the
  first cache in the path that has it.  If the photo is not present in
  any of the caches it is fetched from the backend storage via the
  transcoding tier (4--6).}
\label{fig:download_path}
\end{figure}

\parab{Decompressing on Download.} 
A typical download path for a large-scale photo sharing service in
shown in Figure~\ref{fig:upload_path}.  Users download photos from the
first on-path cache they encounter with the photo: the device cache
(1), the edge cache that handles their request (1-2-8), or the origin
cache (1-2-3-7-8).  If the photo is not in one of those caches it is
fetched from the backend storage system via a transcoder (1--8).

The transcoder converts the photo from the format and size as stored
in the backend to what will be delivered to a user device.  For
instance, Facebook transcodes stored JPEGs into the WebP format before
sending them back to Android devices~\cite{facebook_webp}.  The
transcoder tier also resizes photos depending on their
destination~\cite{huang13fbcdn}.  For instance, a desktop user with a
large open window may receive a larger version of the photo than a
different desktop user with a smaller open window.

The transcoding tier is again a natural place to do decompression on
the download path because it is already manipulating photos. This
placement of decompression on the download path leads to a requirement
that it have low complexity, to ensure both low latency for user
requests for photos and a small impact on the required size of the
transcoding tier. 

\parab{Complexity/Storage Tradeoff.} \sysname trades-off additional complexity for greater
savings in storage. The storage saving is of more important because that the storage increases linearly with time, but the additional complexity does not:
recent measurements of photo sharing services suggest that each image is viewed many times soon after they are uploaded, and very rarely thereafter~\cite{f4}. This means that the complexity cost is never proportional to time.

\parab{Benefits of Low Complexity Decompression.} Even though the storage saving is more significant over time, it is still important to have low complexity, especially for decompression. The reason is that \emph{low complexity in the download
  path ensures low access latency}.  Large-scale content delivery
services optimize latency aggressively and moderate increases in
latency can negatively impact user experience.


The latency introduced by decompression may not affect all downloads,
because of caching. For example, if an image is decompressed and
transcoded once, it can be cached either at the origin cache or the
edge cache for subsequent accesses (unless the subsequent accesses are
from devices of a different type or a different resolution, in which
case decompression must happen again). 
However, even if only cache misses require decompression
(about 29\%~\cite{huang13fbcdn}), it is still important to have
a low latency decompression. 

The most well-known efficient JPEG recompression scheme, PackJPG~\cite{packjpg}, can provide about 20\% additional compression
in photo sizes, but has significant decompression
complexity.\footnote{A more recent recompression scheme, Lepton~\cite{lepton}, does achieve lower latency than PackJPG. We will evaluate Lepton in Section~\ref{eval}.} For downloading a 2048$\times$1536 image from the backend, PackJPG's recompression inflates the
latency of the fastest 40\% of downloads by more
than 50\%. Even though the decompression latency is incurred only for
cache misses, we still see significant impact on the overall latency
distribution (from all the layers of the cache stack):
\tld80\% of the distribution incurs more than 150ms additional
latency.\footnote{The methodology for this experiment is
  explained in Section~\ref{collateral}.}  For commercial photo sharing services, these increases in latency may not be acceptable.

These photo services can also reap other benefits of recompression.
For example, moving the
decompression and transcoding close to the clients can have two
important benefits.  First, caches would now be more storage
efficient, because they would store recompressed versions of the
photos. This would result in fewer accesses to the backend storage
and caches. Second, the bandwidth required between the caches and the
backend storage would be reduced because only recompressed images
would need to be transferred. However, to reap these benefits, the
additional decompression latency would affect \emph{all the
  accesses}. In this case, low complexity decompression is even more
important, using the
state-of-the-art PackJPG, with its high decompression complexity, can
increase latency by more than \emph{half a second}.

For this reason, we make low complexity decompression a primary
design requirement for \sysname and \lossyname.

%% file: recompression.tex
\section{Low-Complexity Recompression for Photo Sharing Services}
\label{sec:romp-lromp}
\label{sec:recompression}

In this section, we discuss the need for a new recompression strategy, and then describe the design of ROMP and L-ROMP, which recompress JPEG images to reduce the storage requirements of large-scale photo sharing services significantly with low complexity overhead.

\subsection{Why a New Recompression Strategy?}

\parab{Limitations of Traditional Approaches.} While photo storage services such as Flickr maintain a copy of the
original uploaded images, large-scale photo sharing services modify
the uploaded image in order to manage the scale of their storage
infrastructure. Specifically, Facebook (a) resizes images and (b)
recompresses them using a smaller quality parameter.

Resizing, while useful in managing storage, has its limits. Camera
resolution has been increasingly steadily over the last five years, as
have display resolutions, even on mobile devices. As a result, it is
likely that photo sharing services will increasingly face pressure to
serve high-resolution images in the near future, so photo sharing
services will need additional tools to manage photo storage. 

Recompressing an image by reducing JPEG's quality parameter (requantizing) is a
convenient knob for managing storage; for example, most cameras
generate high-quality images at quality parameter of 95, but Facebook reduces
the quality parameter on an uploaded image to \tld75. To achieve further storage saving, we have to apply the second quality parameter reduction. As we demonstrate experimentally in Section~\ref{eval}, two consecutive quality parameter reductions lead to unacceptable quality degradation.

In summary, while resizing and quality adjustments are currently used,
they are not a viable solution for producing additional storage
savings. Thus, in this paper we explore a completely different
approach for recompression, enabled by the unique setting of photo-sharing services.

\parab{New Opportunities in Photo-Sharing Services.} As opposed to traditional sender/receiver compression scheme where sender encodes the image and the receiver decodes it, the large-scale photo sharing services represent a new compression setting, with some special properties that offer opportunities for better compression.\footnote{We use encode/decode and compress/decompress interchangeably.}

\parae{Property 1: \emph{Collocated} Encoder/Decoder.}\label{the-collocated-encoderdecoder-setting} Instead of having two distinct sender/receiver entities where
the encoder and the decoder are in physically separate locations, the recompression setting in photo-sharing service includes both compression and decompression \emph{within the service}. In traditional distinct encoder/decoder setting data is typically encoded in one
of two ways. One is using a default codec (\emph{e.g.}, JPEG Standard)
that is present at the sender and receiver of the data and does not need
to be transmitted with it. The other is using a customized codec table
(\emph{e.g.}, JPEG Optimized) that is customized at the sender,
transmitted along with the data to the receiver, and then used to decode
the data. A customized codec table can typically represent the data more
compactly than a default table, but there is a tension between the size
of the codec table and the compactness of its representations. Larger
customized codec tables often lead to higher compression because they
can model the data more effectively. Yet, the customized codec table
must be kept small to ensure that the space savings obtained by the
richer encoding is not negated by having to transmit the table along
with the data.

In the collocated setting the encoder and
decoder are at the same location and thus the codec table used for
encoding the data can be decoupled from the data itself. This decoupling
removes the constraint on the size of codec tables: the codec tables can
be much larger than what is practical for individual files because they
are shared across many files and stored separately. As a result, this new design can potentially
reduce the storage of a service beyond what standard
approaches can achieve.

\begin{figure}
\centering
\noindent\includegraphics[width=0.95\linewidth]{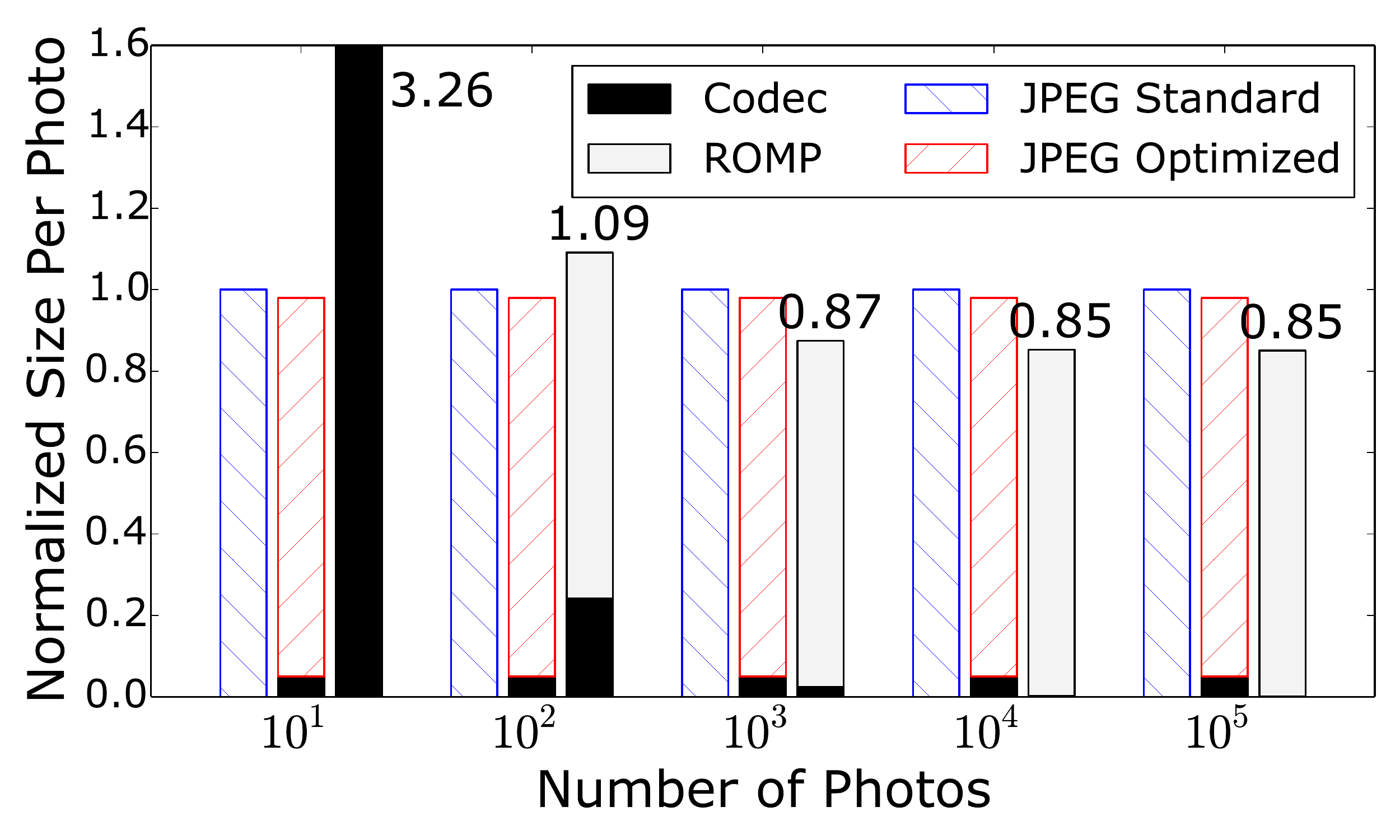}
\caption{Normalized storage size per photo of codec + image using JPEG Standard, JPEG Optimized, and ROMP for an increasing quantity of $2048\times1536$ photos. ROMP makes sense at the large scale it was designed for where the size of its larger codec is negligible compared to the storage saving it offers.}
\label{fig:saving_vs_photos}
\end{figure}

\parae{Property 2: \emph{Large-Scale} Photo Storage.}\label{large-scale-storage} Despite the freedom to use large codec tables, the codec table
constitutes storage overhead that must be carefully managed. However,
the large-scale aspect of photo-sharing services helps in this regard: the same codec
overhead that would negate the additional storage reduction in a
small-scale case will become negligible for large-scale case as it will
be amortized across the entire storage (Figure
\ref{fig:saving_vs_photos}) because the codec tables are shared across
many photos. At small scales, this large codec tables (Section \ref{romp})
can be much larger than the image itself. In contrast, at large scale,
the size of the ROMP codec is negligible compared to the size of the
stored images, \emph{e.g.}, in ROMP design, it needs to store 10,000 images to
make the codec negligible. Our evaluation in Section \ref{eval} shows
that these larger shared tables can compress photos better than
traditional customized tables, and have other benefits across the photo
stack.

In summary, enabled by the collocating of encoder/decoder and the
large-scale setting, large codec tables can be used which would not be practical for traditional small-scale sender/receiver setting. In the following two subsections we describe: (a) ROMP, a lossless coding
technique that uses enhanced Huffman tables, and (b) L-ROMP, a perceptually
near-lossless coder that does not suffer from the significant quality
degradation inherent in re-quantization. At a high level, \sysname and \lossyname make use of a large set of codec tables generated from a large corpus of images, 
where each of the tables is optimized for a specific context, and has similar structure as that of a typical JPEG entropy coding table. Other than using the new codec tables, \sysname and \lossyname proceed block by block and does not involve any transformations or re-orderings of DCT coefficients. This ensures that the coding complexity is very low (see Fig.~\ref{fig:sysname}), which is approximately equivalent to a JPEG entropy decoding 
followed by a JPEG entropy coding. Together, these techniques
can reduce photo storage requirements by 25\% or more, a significant
gain for large photo sharing services that store billions of images.

\subsection{ROMP: Lossless Coding using Enhanced Huffman Tables}
\label{sec:codec}\label{romp}

Our design, \sysname, exploits 
the decreasing cost of storing and managing large tables 
by designing \emph{context-sensitive coding tables} that result in
lossless compression. Recall that JPEG's Huffman tables are used to code symbols (information
about run-lengths and quantized values), 
based on the expected frequency of symbols' occurrence.
\sysname learns \emph{context-sensitive} Huffman tables by
learning the empirical probability of occurrence of these symbols from a large
corpus of images. This learning leverages the availability of such
corpora in a large-scale photo service. Concretely, \sysname uses below two
insights derived from properties of natural images to obtain context-sensitive Huffman tables.

\begin{figure}[t]
    \centering
    \includegraphics[width=0.75\columnwidth]{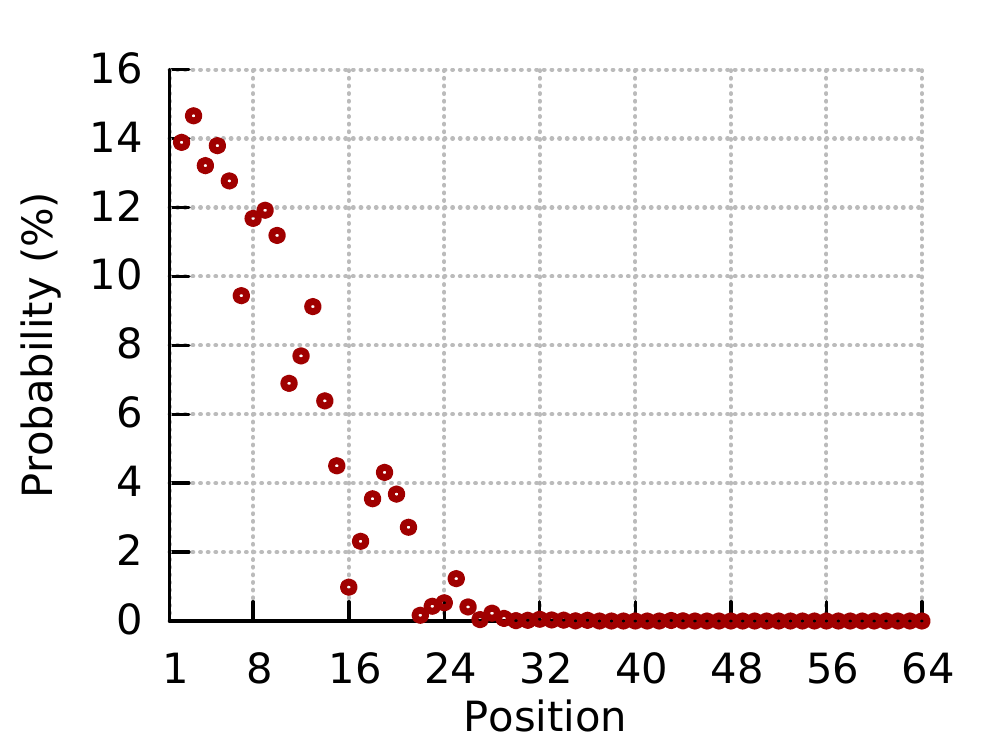}
    \shortenfigurecaption
    \caption{Probabilities of 3 Bit Coefficients at different positions.}
    \label{fig:position_distribution}
    \vspace{-0.25in}
\end{figure}

\parab{Position-dependence.}
The empirical probability of
occurrence of a symbol can depend on its position in the 
zig-zag scan. 
That is, for a given symbol, its 
empirical probability of occurrence  at position $p_a$ along the 
zig-zag scan is likely
to be different from its empirical distribution at position $p_b$
(Figure~\ref{fig:position_distribution} illustrates this). 
For example, for natural images, it is known that
non-zero coefficients are increasingly unlikely at higher
frequencies~\cite{lam2000DCT}. Thus, if $p_b > p_a$ then a non-zero 
value will be less likely at $p_b$. For this reason, \sysname generates
different tables for different positions: i.e., position is one aspect
of the context used for encoding. Thus, the same symbol may be encoded
using different bit-patterns depending on the position where it occurs. 

\parab{Energy-dependence.}
The second insight is based on the observation that, relative to image
sizes predominantly in use today, an $8\times8$ block represents a
very small patch of the image.  To see this, imagine capturing the
same visual information (a photo of a person, say) with two cameras of
different resolutions, and then using JPEG encoding for both
images. Clearly $8\times8$ blocks in the higher resolution image
represent smaller regions of the field of view, and thus will tend to
be smoother.  This has two useful implications. First, information
within a block will tend to be smooth, with additional smoothness
expected for larger images. Smoother images are such that coefficients
at higher frequencies tend to be smaller. Second, because the same
region in the field of view comprises more blocks when larger images
(higher resolutions) are generated, it becomes more likely that
neighboring blocks will have similar characteristics.

We exploit these two ideas by creating tables such that the
probability of occurrence of a symbol at a given position can also
depend upon the \emph{energy} of other coefficients within the
block (\emph{intra-block} energy) and of neighboring blocks
(\emph{inter-block} energy). For a given \syma
that occurs at zigzag position $p$ of the $n$-th block, we use the
average of the observed coefficient sizes in a block as an estimate of intra-block 
energy: 
\begin{equation}
intra(n,p)=\frac{1}{p-1}\sum_{i=1}^{p-1}\frac{SIZE(b_n(i))}{max_{SIZE}(i)}
\label{formula:intra}
\end{equation}
where $b_{n}$ denotes the $n$-th block, and
$b_n(i)$ denotes the coefficient at position $i$, 
$SIZE(\cdot)$ denotes the bits required to represent the
amplitude of the coefficient, $max_{SIZE}(i)$ is the observed maximum
coefficient size for position $i$ of images in the training set.

Similarly, the inter-block energy value is estimated based on the 
average sizes of coefficients in nearby blocks: $F$
nearby zigzag positions in $B$ adjacent prior blocks (ROMP uses $F=5$ and $B=3$):
\vspace{-0.3cm}
\begin{equation}
inter(n,p)=\frac{1}{B\cdot F}\sum_{i=n-B}^{n-1}\sum_{j=p}^{p+F-1}\frac{SIZE(b_i(j))}{max_{SIZE}(j)}
\label{formula:inter}
\end{equation}

\begin{figure}[t]
  \centering\includegraphics[width=0.65\linewidth]{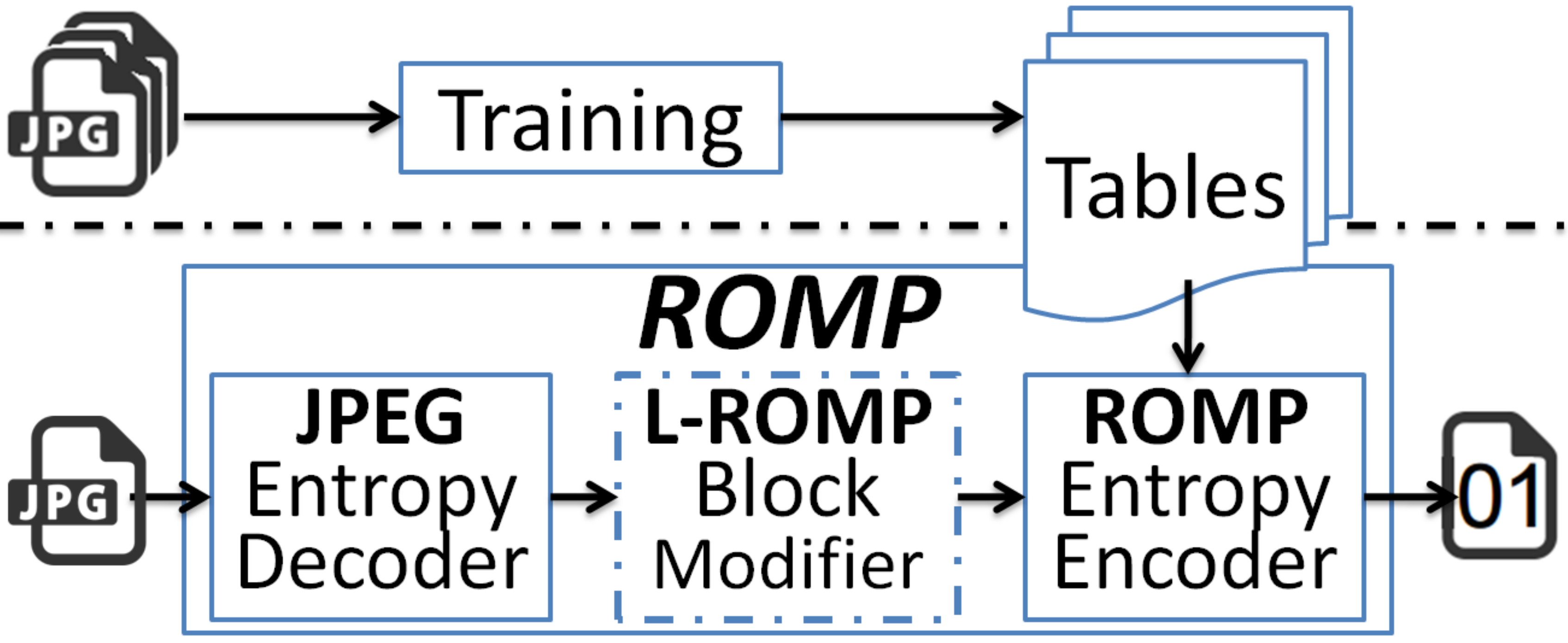}
  \caption{\sysname Encoding Architecture}
 \label{fig:sysname}
 \vspace{-0.25in}
\end{figure}

\parab{Putting it all together.} Context in \sysname is defined by a triple $<p,i,e>$ to define context: zigzag position $p$, intra-block energy $i$ and inter-block energy $e$. Note that the statistical dependencies captured by this context information 
are well known in image coding and exploited by 
state of the art compression techniques, but to the best of our knowledge we are the first to take advantage of them for low latency JPEG transcoding (where complexity is minimized by using memory to store many, context-dependent Huffman tables). 

At a high-level, \sysname
works as follows (Figure~\ref{fig:sysname}). From a training set of images, \sysname learns a Huffman table for each unique context (i.e., for each unique combination of position, intra- and inter-block energy). For \syma that occurs in any image in the training set, \sysname first determines its context triple and then gathers it together with other \symas belonging to the same context. After gathering all the \symas for each context, \sysname can generate a table for this context based on the number of occurrences of each. \sysname pre-defines 20 different energy levels for both intra-energy and inter-energy, which leads to \tld$64\times20\times20=25600$ different contexts and Huffman tables to be learned.\footnote{These tables take up  less than 16MBs of the memory, a negligible memory usage increment for modern machines.} These tables are quite different, which allows \sysname to achieve better compression over standard JPEG. When an image is uploaded, \sysname decodes the JPEG image, computes the corresponding triple $<p,i,e>$, the then uses the learned Huffman tables to re-code the image. Before delivering an image to a client, \sysname reverses its context-sensitive entropy code, then applies the default JPEG entropy code.

\subsection{L-ROMP: A Gracefully Lossy Coder}
\label{sec:codec_lossy}

\sysname's entropy coder is lossless with respect to the uploaded
JPEG. In this section, we describe \lossyname, which introduces a controlled, perceptually insignificant, amount of loss
(or \emph{distortion}) in uploaded images, as a way of achieving
further savings in photo storage. 

\begin{figure}[t]
  \centering
    \includegraphics[width=1.05\linewidth]{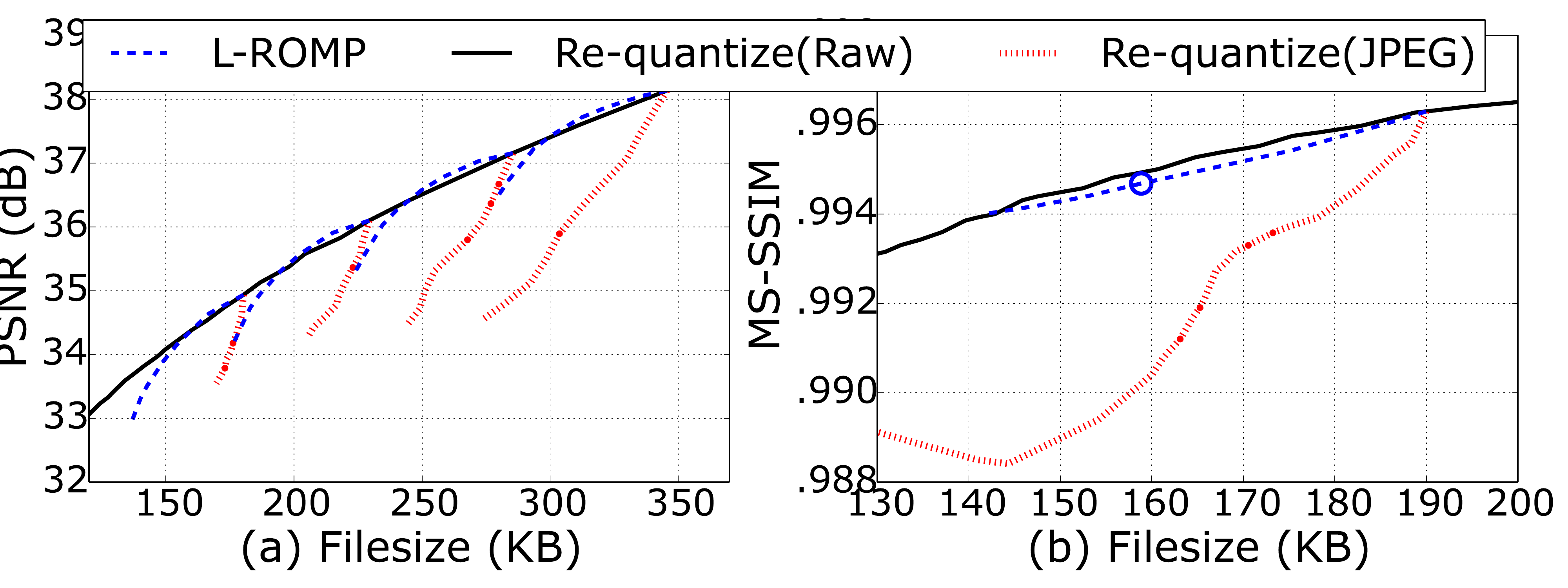}
\vspace{-0.5cm}
  \caption{Rate/distortion performance of \lossyname, compared to re-quantization from raw image and from JPEG image using Tecnick images of 1200$\times$1200. \textbf{(a)}: Using PSNR as the quality metric, and showing performance on JPEG images of four quality parameters (70,80,86,90). \textbf{(b)}: Using MS-SSIM metric, and focusing on JPEG images of quality paramter of 75; ``o'' marks the perceptually lossless setting of \lossyname.}
 \label{fig:better_transcoding}
\vspace{-0.5cm} 
\end{figure}

As discussed previously, users upload high quality JPEG images and  many photo sharing services, e.g., Facebook, change the JPEG quality  parameter to a lower level in order to ensure predictable storage usage, a step that introduces additional distortion. Figure~\ref{fig:better_transcoding} quantifies the dramatic increase in distortion caused by re-quantization, compared to quantizing the original raw image directly to the target quality parameter, showing rate-distortion performance with two different objective quality metrics: PSNR and MS-SSIM. The ``Re-quantize (raw)'' curve is obtained by reducing the quality parameter when compressing the original raw images, while the ``Re-quantize (JPEG)'' curve is obtained by re-quantizing a high-quality JPEG image derived from the same set of raw images. The former curve represents much more graceful degradation. By contrast, re-quantization introduces much more distortion for the same size reduction. This happens because the latter suffers from cumulative errors due to two consecutive re-quantizations.

\lossyname avoids re-quantizing coefficients, but
introduces distortion by carefully setting \emph{some} non-zero
(quantized) coefficients to zero, a specific instance of a general
idea called thresholding~\cite{thresholding}. The intuition behind thresholding is
that, by setting a well-chosen non-zero coefficient to zero, it is
possible to increase the number of consecutive zero-valued
coefficients in the sequence of coefficients along the zig-zag scan. 
This in general helps 
reduce size as it replaces two separate runs of zeros together 
and a non-zero coefficient value, by a single run of zeros. Optimization of coefficient thresholding has been considered from a 
rate-distortion perspective in the literature~\cite{ortega1998rate,thresholding}, we are not
aware of it being explored as an alternative to re-quantization in
large photo sharing services.
Here we use a simplified version where only coefficients of size equal to 1 (i.e., $1$ or
$-1$) can be removed. This means that the distortion increase for any coefficient 
being removed will be the same.\footnote{Note that the actual MSE will be different if 
two coefficients have same value $1$, but different frequency weights in the quantization 
matrix. However, by ignoring this difference we take into account the different perceptual 
weighting given to each frequency and obtain better perceptual quality.} Thus, we only need to decide if for a given 
coefficient the bit-rate savings are sufficient to remove it. We make the decision by introducing a \emph{rate threshold} and only thresholding a coefficient if the bits saving by doing so would exceed this threshold.

However, setting too many coefficients to zero within a block can
introduce local artifacts (e.g., blocking). Thus, \lossyname uses 
a \emph{perceptual threshold} $T_p$ that limits 
the percentage of non-zero  
coefficients that will be set to zero.
By doing this \lossyname can guarantee that the block-wise SSIM with respect to the original 
JPEG is always higher than $1-\frac{T_p}{2-T_p}$. 
For example, if we use $T_p=0.1$ (i.e., we can threshold at most 10\% of the non-zero coefficients), then block-wise SSIM metric is guaranteed to be higher than 0.947. 
The proof of such bound is based on results from~\cite{ssim_distortion} and is omitted for brevity.

In Figure~\ref{fig:better_transcoding} (a), we observe that \lossyname degrades PSNR more gracefully than simply re-quantizing. We see that with conservative thresholds, \lossyname's curve is actually higher than re-quantizing from the raw image curve, illustrating the efficiency of \lossyname's trading distortion for bits-saving. \lossyname performs equally well on MS-SSIM metric (plot (b)). To achieve perceptually lossless compression, we also conducted a subjective evaluation, by developing a comparison tool that can choose thresholds and shows the image at the chosen setting as well as size reduction. We find that using rate threshold of 2.0 and perceptual threshold of 0.4 provides maximum bits-saving without noticeable quality distortion (``o'' of Figure~\ref{fig:better_transcoding} (b)). We use these parameters for L-ROMP.

Finally, \lossyname can be easily introduced into \sysname's
pipeline: before applying the
context-sensitive entropy coding, 
\lossyname's thresholding can be
applied to each block (Figure~\ref{fig:sysname}). No changes are
required to \sysname's entropy coder.

\subsection{Encoding/Decoding in Parallel to Reduce Latency}

A recompression codec that can achieve both high compression and low coding latency is ideal to photo sharing services. However, generally speaking, a high compression ratio codec introduces higher complexity, and thus higher coding latency at the same time. With reducing ``complexity'', a parallelized codec can reduce the coding latency, and some codecs have explored this idea~\cite{lepton}. Parallelizing can reduce the latency in terms of time, but because the codec still requires the same number of CPU cycles, it cannot improve the coding throughput. It means that, when CPU resources is the bottleneck, parallelized codecs would not have benefits.\footnote{Actually, parallelized codec will introduce additional CPU overhead to enable parallelism.} Because of this, the number of CPU cycles is the right metric for complexity. 

However, when CPU resources is not the bottleneck, the capability of enabling parallelism is an important feature for recompression codecs to further reduce coding latency. \sysname can be easily extended to parallelized, multi-threaded version. \sysname just needs to break the original image into $N$ sub-images to enable \sysname's encoding in $N$ threads. \sysname needs to save each encoded sub-image separately, with this sub-image's offset of the original JPEG image, i.e., for the original JPEG, where is the first bit of this sub-image. At the decoding, \sysname can enable $N$ decoding threads, each decodes one encoded/compressed sub-image, but writes the decoded bits to the right location indicated by the offset information. By doing so, these $N$ decoding threads collectively recover the original JPEG image.

Enabling $N$ threads can take the encoding/decoding latency to approximately $1/N$ of the original, single-thread \sysname. The next question is whether the enabling of parallelism negatively affects compression. Multi-threaded \sysname requires more bits to represent the original JPEG, for two reasons. The first reason is that, multi-threaded \sysname needs to store the extra offset information. The second reason is that, to encode one block, \sysname uses $B$ adjacent prior blocks to predict current block (inter-block energy-dependence). But for the first $B$ blocks of each image, this information is not complete. This affects the predictability, and then the compressibility. Fortunately, these two penalties are both negligible for \sysname's compression. Offset information requires 64 bits per thread for an image that is no larger than 4GB. This is clearly negligible, as each thread usually handles hundreds of KBs data, a less than 0.01\% penalty. The second penalty is also insignificant. Each thread will encode thousands of blocks (e.g., a 2048$\times$1536 image contains ~50000 blocks), only 3 (\sysname uses $B=53$) of them are lack of information for prediction is not a big deal. On the other hand, for \lossyname, enabling parallelism would not introduce any penalties. In our experiment, a 4-thread \sysname reduces the coding latency to less than 1/3 of the original single-thread \sysname's latency, while only introduces 0.01\% of extra bits. We conclude that \sysname and \lossyname can both be extended to multi-threaded version easily.

%% file: eval.tex
\section{Evaluation}\label{eval}

We experimentally explore three key questions:

\begin{itemize}
\itemsep1pt\parskip0pt\parsep0pt
\item
  How do ROMP and L-ROMP compare to the state of the art
  in terms of compression ratio and complexity under a variety of
  settings?
\item
  What compression can ROMP and L-ROMP achieve and what storage savings does that translate into?
\item
  What collateral benefits does ROMP and L-ROMP enable?
\end{itemize}

\subsection{Methodology}\label{methodology}

\paragraph{Implementation}\label{implementation}

Our evaluations use an implementation of ROMP that has two software
components: a training script and a codec. The training script is
implemented in Python and takes a training set of JPEG images, decodes
them, and generates Huffman tables. Training is done once and is
off-path for photo uploads and downloads and so it is not included in
complexity measurements. The codec is implemented on top of
\texttt{libjpeg-turbo} \cite{libjpegturbo}. The implementation of
L-ROMP is an extension of ROMP that adds an additional
block-modification stage to the image processing pipeline. ROMP and
L-ROMP are publicly available \cite{romp}.

\paragraph{Baselines}\label{baselines}

To quantify the compression ratios and complexity of ROMP and L-ROMP, we
compare them with lossless JPEG codecs and alternative photo formats.
The lossless JPEG codecs include all that we are aware of that have
publicly available implementations: JPEG Standard, JPEG Optimized, JPEG
Progressive, JPEG Arithmetic, MozJPEG \cite{mozjpeg}, PackJPG
\cite{packjpg} and Lepton~\cite{lepton}. The alternative photo formats include WebP and
JPEG2000.

\paragraph{Image Sets}\label{image-sets}

Our evaluation uses two sets of images:

\begin{itemize}
\itemsep1pt\parskip0pt\parsep0pt
\item
  \emph{Tecnick} \cite{tecnick} is an image set of 100 images, each
  available in many resolutions up to $1200\times1200$. The images are
  in a raw format (PNG), which allows us to transcode them to JPEGs of
  different resolutions and different quality parameters. We use Tecnick
  because it is commonly used in image processing benchmarks.
\item
  \emph{FiveK} \cite{fivek} is an image set from MIT-Adobe that
  contains 5,000 publicly available raw images taken from different SLR
  cameras. This data set is used to evaluate images with higher
  resolution than the max of $1200\times1200$ in the Tecnick image set.
\end{itemize}

\paragraph{Metrics}\label{metrics}

We evaluate our compression schemes and their baselines on two metrics:
compression ratio, and encoding/decoding time. \emph{Compression ratio}
measures the storage efficiency and is the ratio of saved storage over
old storage. More precisely, let $s'$ be the size of an image generated
by a scheme and $s$ be the size generated by JPEG Standard. Then, the
compression ratio is $\frac{s-s'}{s}$. The \emph{encoding time} is the
time to recompress from JPEG Standard and the \emph{decoding time} is
the time to decompress back to JPEG Standard. To make fair coding complexity comparison among schemes, we measure the encoding/decoding time of single-thread version of each scheme.\footnote{The only scheme that supports parrellelism is Lepton~\cite{lepton}, we use its single-thread version for evaluation.}

\paragraph{Testing}\label{testing}

The images used for our experiments are obtained from the image sets
described above. For the small Tecnick image set, we use 10-fold cross
validation: the 100 image set is divided into 10 groups of 10 images; we
test on one group of 10 images after training on the other 90 images,
and repeat this procedure for every group in order to test on every
image of the set. For the FiveK image sets we use 1000 randomly chosen
images for training and the rest of the images for testing.

\subsection{Compression \& Complexity}\label{compression}

\sysname and \lossyname are designed to provide high compression ratios
with low complexity. This subsection compares the complexity and
compression ratios of ROMP and L-ROMP against the state of the art. We
evaluate under a variety of settings with an eye to the future where
photos will be larger in size and higher in resolution.

\begin{figure*}
\begin{minipage}{0.333\textwidth}
\centering
\begin{subfigure}{1\linewidth}
  \centering
  \includegraphics[width=1\linewidth]{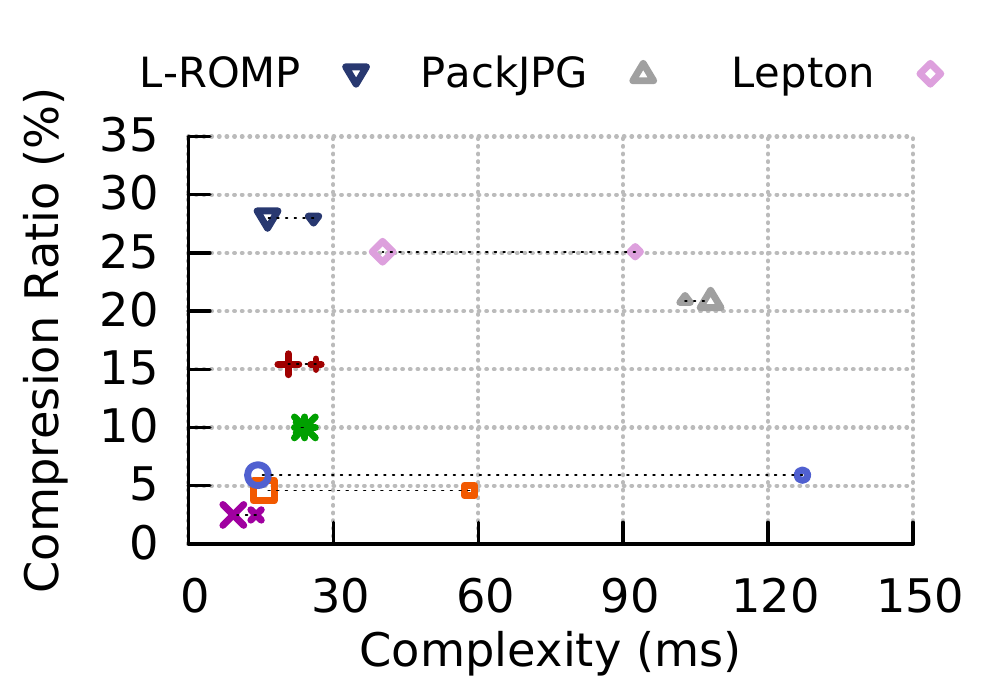}
  \caption{1152$\times$864, quality is 75.}
  \label{fig:fivek_tradeoff-1}
\end{subfigure}\\
\begin{subfigure}{1\linewidth}
  \centering
  \includegraphics[width=1\linewidth]{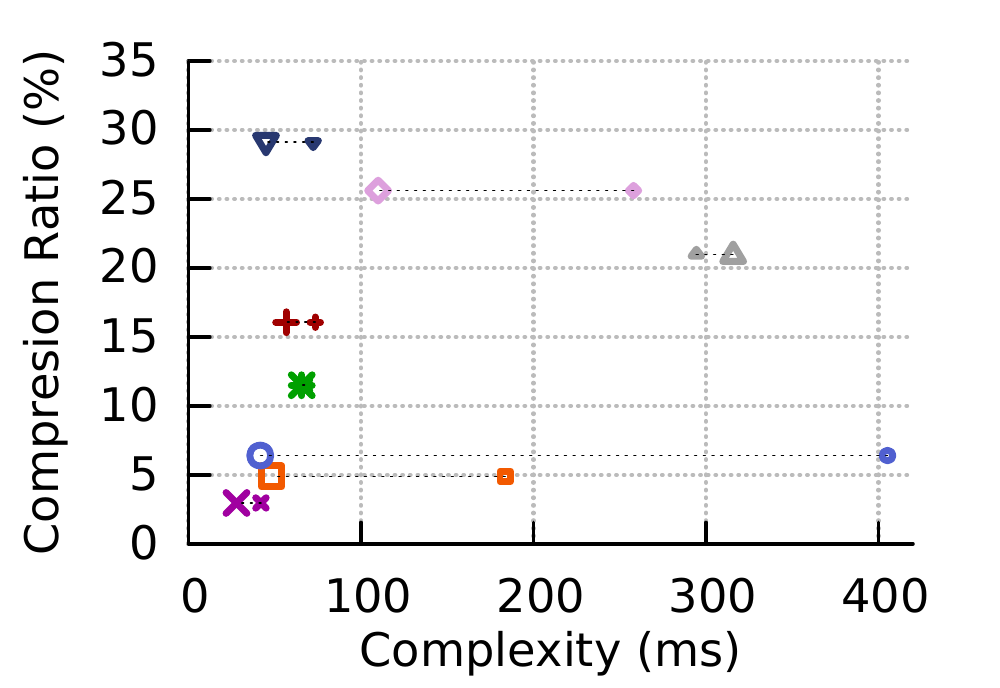}
  \caption{$2048\times1536$, quality is 75.}
  \label{fig:fivek_tradeoff}
\end{subfigure}
\caption{Compression/complexity tradeoff of FiveK image set. The bigger marker indicates the decoding complexity while the smaller shows the encoding complexity.}
\end{minipage}
\begin{minipage}{0.333\textwidth}
\centering
\begin{subfigure}{1\linewidth}
  \centering
  \includegraphics[width=1\linewidth]{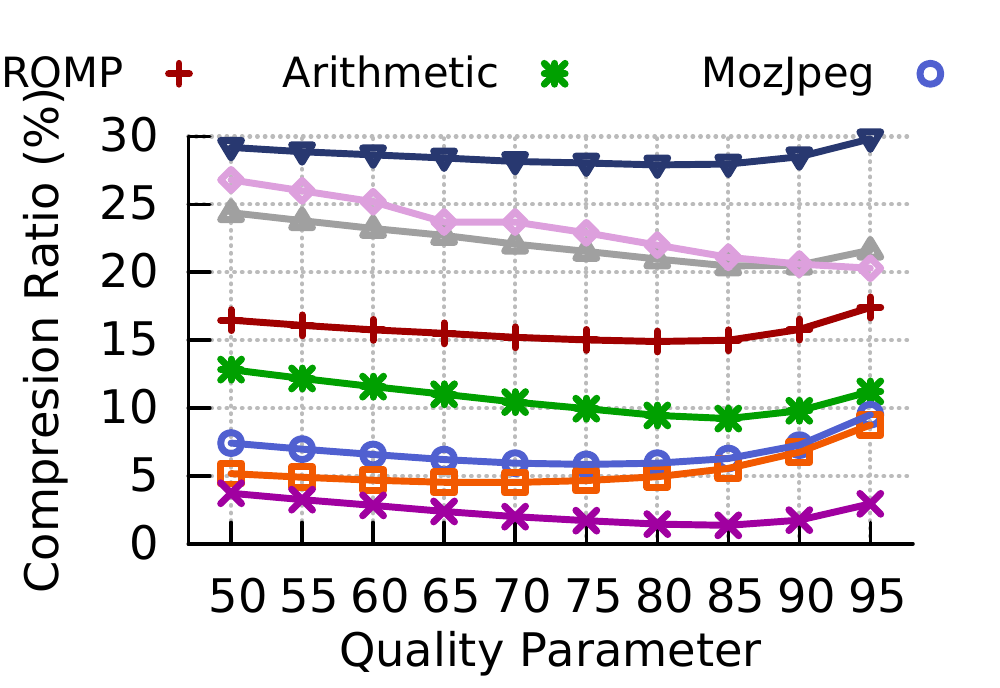}
  \caption{$1200\times1200$}
  \label{fig:tecnick_quality}
\end{subfigure}\\
\begin{subfigure}{1\linewidth}
  \centering
  \includegraphics[width=1\linewidth]{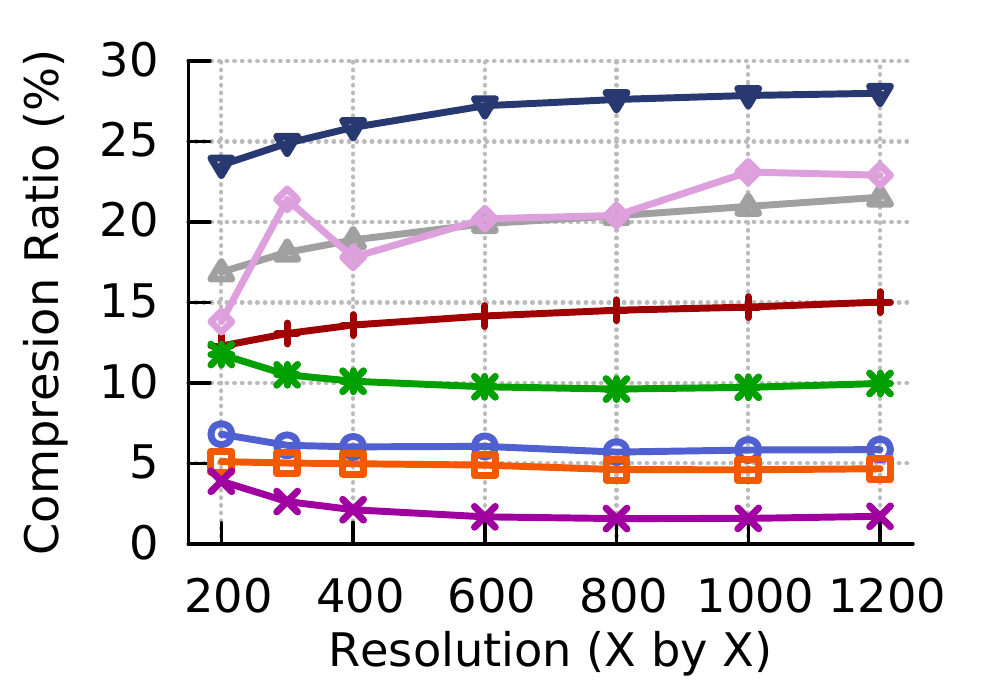}
\caption{Quality is 75}
\label{fig:tecnick_filesize}
\end{subfigure}
\caption{The compression ratio over JPEG Standard baseline as (a) quality parameter changes and (b) resolution changes for Tecnick image set.}
\end{minipage}
\begin{minipage}{0.333\textwidth}
\centering
\begin{subfigure}{1\linewidth}
  \centering
  \includegraphics[width=1\linewidth]{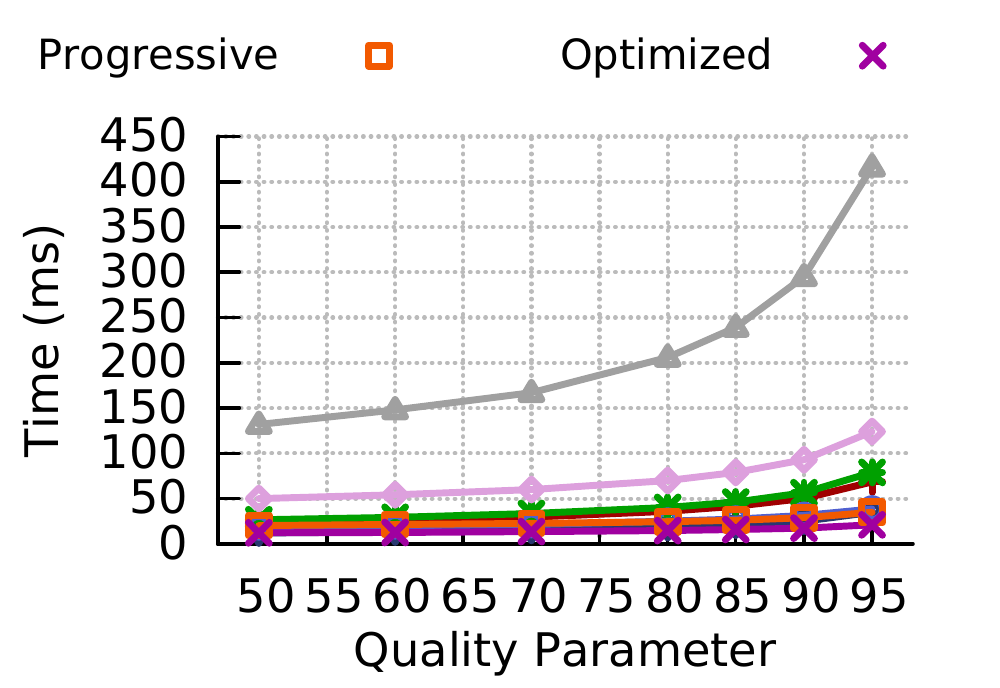}
  \caption{$1200\times1200$} 
  \label{fig:complexity_quality}
\end{subfigure}\\
\begin{subfigure}{1\linewidth}
  \centering
  \includegraphics[width=1\linewidth]{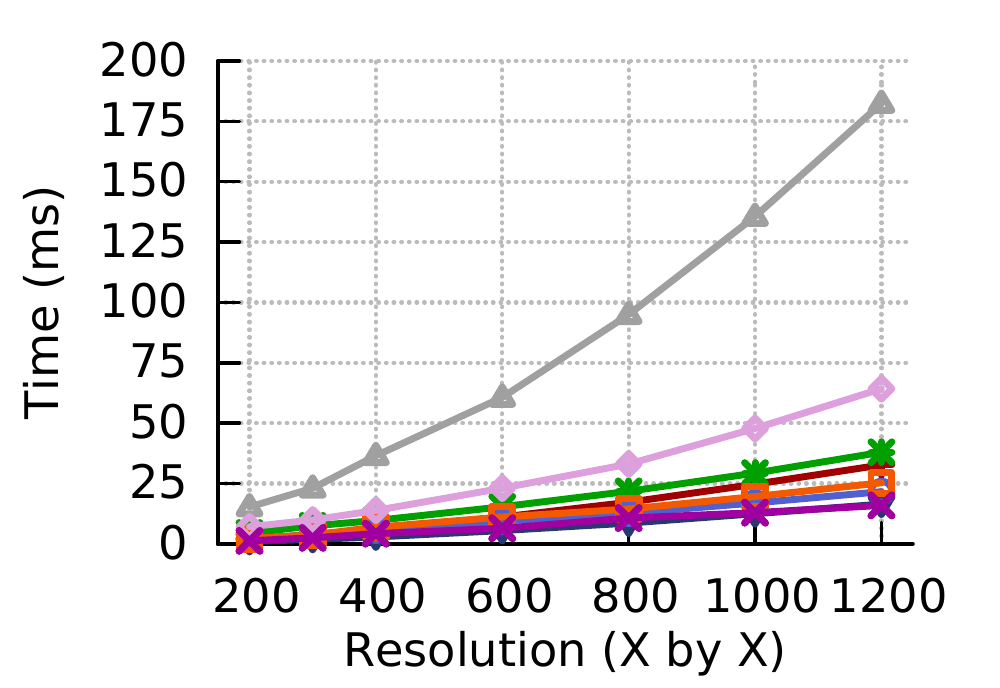}
  \caption{Quality is 75}
  \label{fig:complexity_coding}
\end{subfigure}
\caption{The decoding complexity as (a) quality parameter changes and (b) resolution changes for Tecnick image set. (Y-axis ranges differ for readability.)}
\end{minipage}
 
\end{figure*}

\paragraph{Compression/Complexity
Tradeoff}\label{compressioncomplexity-tradeoff}

Compression techniques represent different points in the complexity
vs.~compression tradeoff. In general, with higher
complexity----\emph{i.e.}, higher encoding/decoding time---higher
compression ratio are possible. Figures \ref{fig:fivek_tradeoff-1} and
\ref{fig:fivek_tradeoff} present this tradeoff for the alternatives
considered in the paper, for two different image resolutions and a
quality parameter of 75 on the FiveK
dataset. We exclude results for transcoding back and forth from other image
formats, i.e., WebP and JPEG2000, because they are clear outliers: they
require 700--2900ms to decompress a $2048\times1536$ photo.

For encoding time, ROMP is comparable to JPEG Arithmetic, and much
faster than other competitors. In particular, the high compression schemes, PackJPG and Lepton, both require roughly 4$\times$ encoding time. Compared to ROMP, L-ROMP's additional
step of thresholding does not induce any extra complexity. Decoding time
is the more relevant metric for ROMP because it affects user-perceived
delay. ROMP's decoder is slightly faster than JPEG Arithmetic,
comparable to JPEG Progressive and MozJPEG, 5$\times$ faster than
PackJPG and 2$\times$ faster than Lepton. L-ROMP's decoder is identical to ROMP, but after thresholding
the image becomes smaller, which makes it \tld20\% faster than ROMP.
Interestingly, PackJPG is the only scheme whose decoding time is higher
than its encoding time. Finally, ROMP has higher compression ratio than
almost any other alternative: only PackJPG and Lepton achieve higher compression ratio.

This experiment shows that ROMP and L-ROMP occupy a unique position in
the tradeoff space, achieving both high compression ratio ($15-29$\%)
and low complexity ($60$ms encoding/decoding time for a
$2048\times1536$ image). By contrast, the other high-compression schemes,
PackJPG and Lepton, have a compression ratio of 20\%, but this comes at
considerable complexity cost. Their encoding time are over 250ms, and decoding time are over 110ms (more precisely, PackJPG's decoding is more than 310ms), which we consider
unacceptable because it would shift the latency distribution
significantly (Figure \ref{fig:collateral_benefits_latency}).

\paragraph{Compression Ratio as Quality and Resolution
Increase}\label{compression-ratio-as-quality-and-resolution-increase}

As camera technology continues to improve, we expect users to upload
images with higher quality and resolution and thus the compression
performance for higher quality parameters and resolutions is important.
Figure \ref{fig:tecnick_quality} evaluates the compression ratio of
ROMP, L-ROMP, and baselines as a function of varying quality parameter
on the Tecnick dataset. These schemes are generally robust to changing
quality factors, but with their lowest compression ratio around quality
parameter 75. We believe this is because we are comparing against JPEG
Standard and its Huffman tables are optimized for the widely used quality
parameter, 75. The robustness of ROMP's compression is validated by this
experiment, where we see compression ratios over 15\% for all quality
parameters.

Figure \ref{fig:tecnick_filesize} explores the effect of the trend
towards higher resolution images. The figure shows the compression ratio
of ROMP and L-ROMP over JPEG Standard for increasing image resolutions.
ROMP's compression ratio increases with image resolution, in contrast
with all other low complexity alternatives. ROMP's compression ratio is close to PackJPG and Lepton at the highest quality parameter. This is an important
property given the trend towards larger image sizes with higher quality parameters. One reason for this good property is that ROMP can train and use different coding tables for different image parameters, while other schemes might have poor performances on certain image parameters. We have also
verified this trend in the FiveK dataset (omitted for brevity).

\paragraph{Decoding Complexity as Quality and Resolution
Increase}\label{decoding-complexity-as-quality-and-resolution-increase}

Figure \ref{fig:complexity_quality} shows how decoding time scales with
increasing quality of images. We observe that schemes with low decoding
complexity scale well; the decoding time for PackJPG and Lepton, however, scales
poorly with image size and image quality. Figure
\ref{fig:complexity_coding} shows how decoding time scales with
increasing resolution of images. We see a similar trend to increasing
quality; low decoding complexity schemes scale well. This re-affirms our
findings that ROMP and L-ROMP are better codecs.
ROMP occupies a sweet-spot in the complexity/compression tradeoff space:
even though PackJPG and Lepton have higher compression ratio, it scales poorly with
the trend towards larger, higher quality images. L-ROMP is an even
better choice if perceptually indistinguishable changes are acceptable.

\subsection{Storage Reduction for Photo
Backends}\label{storage-reduction-for-photo-backends}

This section estimates the compression ratios ROMP and L-ROMP can achieve
for a real photo-sharing service, i.e., Facebook's photo storage system.

Above experiment demonstrates compression ratio over JPEG Standard, we need to translate above compression result to benefits over JPEG Optimize, which is more popular as it is a clear winner over JPEG Standard. We estimate ROMP and L-ROMP would result in 13\% and 26\% compression ratios respectively on JPEG Optimize (instead of 15\% and 28\% on JPEG Standard). This estimate is done by pre-optimize images we use in above experiment to JPEG Optimize and re-run the experiment. Note that, we see slightly different compression ratios on images with different quality parameter or resolution in above experiment, we download Facebook photos, get the average JPEG quality parameter and resolution and use that to pick these compression ratio values.  These are the values
we use later to calculate the storage reduction and collateral benefits
of deploying ROMP and L-ROMP into the photo-sharing service.

\paragraph{Storage Reduction}\label{tt-ltc-and-te-ltc-reduce-storage-footprints}

The storage
reduction from a compression scheme is greater than simply its
compression ratio because photo sharing services replicate images for
fault tolerance and load balancing. This results in a physical image storage that is a multiple of the logical size of the stored
images, \emph{i.e.}, the effective replication factor. Facebook's Haystack \cite{haystack} has an
effective replication factor of 3.6$\times$ and f4 \cite{f4} has an
effective replication factor of 2.1$\times$.

\begin{figure}
\centering
 
\begin{tabular}{llcc}
\toprule[1.5pt]
            &             & {\bf ROMP} & {\bf L-ROMP} \\ Compression & 0\%         & 13\%           & 26\%             \\ 
 
\midrule
Haystack    & 3.6$\times$ & 3.1$\times$    & 2.7$\times$      \\ f4          & 2.1$\times$ & 1.8$\times$    & 1.6$\times$      \\ 
\bottomrule[1.5pt]
\end{tabular}
 
\caption{New effective replication factor if compression schemes
  were deployed based on the compression ratio over JPEG Optimized for   large photos.}
\label{tbl:storage_reduction}
\end{figure}

Figure \ref{tbl:storage_reduction} shows how ROMP and L-ROMP would
reduce the effective replication factor of Haystack and f4. The
difference between the current effective replication factor and the new
factor is the storage reduction due to the deployment of ROMP and L-ROMP.
For instance, if ROMP was deployed on
Haystack it would reduce the storage footprint by $.5\times$ the logical
size of the images it stores, and if L-ROMP was used on Haystack, it
would reduce the storage requirements almost by the size of one complete
copy of the images ($.9\times$). Similarly, if L-ROMP was used on f4 it
would reduce the storage footprint by $.5\times$ the logical size of the
images.

\subsection{Collateral Benefits}\label{collateral}

This subsection quantifies, using a data-driven model-based
approach, the collateral benefits of placing the decoder in
the edge cache. These collateral benefits include a larger effective
cache size, increased hit rates at the caches, reductions in backfill
requests and bytes, and a reduction in external bandwidth when L-ROMP is
used. L-ROMP can impact download latency, which we also quantify.

\paragraph{Methodology: Data-Driven Model-Based
Estimation}\label{methodology-data-driven-model-based-estimation}

We use a data-driven model-based approach to estimate the collateral
benefits with the deployment of ROMP, on the photo stack of a large
provider described in Figure \ref{fig:download_path}. At a high level,
each box in this figure is associated with a distribution of processing
latencies, and each link with a distribution of transfer latencies. In
addition, caches have an associated hit rate and we model cache hits by
assuming that photos have uniform probability of a cache hit, given by
the hit rate.

We combine multiple measurement results of the current photo stack to
parameterize the model. We use measurements from a 2013 Facebook study
\cite{huang13fbcdn} to get the cache hit rates for the model and we
combine latency measurements from \cite{huang13fbcdn} and our recent
measurement study \cite{dbit2015} to obtain the transfer latencies
for the model. We need to update the model in two ways:
first, the processing latency due to decompression of ROMP or L-ROMP
will need to be added to the edge caches; second, the cache hit rates
need to be re-computed, resulting in changes in the percentage of
requests served by each cache layers and the distribution of download
latency. We discuss these changes in the paragraphs below.

\parae{Cache Hit Rates.} In the absence of ROMP, we assume the cache
hit rates of edge caches and origin caches are based on measurement
results from \cite{huang13fbcdn}. ROMP or L-ROMP would reduce the
size of each image stored in a cache, which would allow each cache to
store more images. To compute this \emph{effective cache size increase}
as a result of compression, we use the following model: for a codec
with a compression ratio of $x$\%, the cache size effectively increases
by a factor of $\frac{1}{1-x\%}$. For example, for L-ROMP, this results
in a cache size \tld1.35$\times$ the original size. We use this as the
new cache size to update the cache hit rates at edge caches and origin
caches (denoted by $H_e$ and $H_o$, respectively) based on Figure 10 of
\cite{huang13fbcdn}.

\parae{Fraction of Requests Served by Cache Layers.} A change in the
cache hit rates in any layer would change the percentage of requests
served by edge caches, origin caches and backend (denoted by $S_e$,
$S_o$ and $S_b$, respectively). At edge caches, we have $S_e=H_e$, at
origin caches $S_o =(1-H_e)H_o$ and $S_b=(1-H_e)(1-H_o)$ at the backend.
Different sets of $S_*$ means the re-distributions the load on
different cache layers, and we use these values to analyze the load
changes of different cache layers, changes to internal/external
bandwidth and also the change in the distribution of the download
latency.\footnote{We use $S_*$ to collectively denote $S_e$, $S_o$, and
  $S_b$.}

\parae{Download Latency.} We estimate the impact on download latency of
changes to $S_*$ by getting the distributions of download latencies from
edge caches, origin caches and backend (denoted by $L_e$, $L_o$ and
$L_b$, respectively) and combining them to form one overall distribution
using the new $S_*$. We obtain $L_e$ and $L_b$ from our previous work
\cite{dbit2015}.\footnote{This study measured the distribution of latencies for Facebook (and other photo providers) by downloading small images from several hundred PlanetLab sites and several thousand RIPE Atlas sites. We extend its results to what we would expect for larger images by using the reported time-to-first-byte latency and the transfer rate of the rest of the bytes.  This will slightly over inflate the latency because the transfer rate of the slow start phase in TCP is generally slower than the congestion avoidance phase.  But, we expect this effect to be small and believe our estimates are representative.}
Separately, we get origin cache to backend latency distribution from the
Facebook study (Figure 7 of \cite{huang13fbcdn}) and subtract that
(in a distributional sense) from $L_b$ to get $L_o$. We then update
$L_*$ by taking ROMP or L-ROMP's decompression latency into account and
shift the distribution uniformly. With both updated $S_*$ and $L_*$, we
can estimate the distribution of overall download latency $L$ as
follows. Let $L_*(x)$ be the probability of latency $x$ms of the
distribution $L_*$, we have $L(x) = S_eL_e(x) + S_oL_o(x) + S_bL_b(x)$.
We can then enumerate on $x$ to get the entire distribution of $L$.

\paragraph{Results}\label{results}

We now present the benefits of deploying ROMP.

\begin{figure}
\centering
 
\begin{tabular}{lcc}
\toprule[1.5pt]
                             & \textbf{ROMP} & \textbf{L-ROMP} \\ Compression                  & 13\%       & 26\%         \\ 
 
\midrule
Effective Cache Size         & 1.15$\times$      & 1.35$\times$        \\ Hit Ratio Increase           &            &              \\
\quad Edge Cache                   & 1.1\%      & 2.5\%        \\
 
\quad Origin Cache                 & 1.6\%      & 3.8\%        \\
 
Reduction in Requests to Backend    & 4.9\%      & 11.4\%       \\ Reduction in Bytes Sent to Edge     & 15.2\%     & 30.5\%       \\ Reduction in External Bandwidth & 0.0\%      & 15.5\%       \\ 
 
\bottomrule[1.5pt]
\end{tabular}
 
\caption{Estimated collateral benefits from deploying ROMP in the Facebook photo caching stack.}
\label{tbl:cache_effects}
\end{figure}

\parae{Cache Hit Rates.} Figure \ref{tbl:cache_effects} shows how
deploying ROMP on the Facebook photo caching stack would change the
effective cache sizes and how this would affect cache hit rates. For
instance, using L-ROMP would result in a 2.5\% increase in the hit rate
at the edge caches, and a nearly 4\% increase in hit rate at the origin
caches. In turn, these increases in the hit ratio can (1) reduce the
number of requests to the storage backend, (2) reduce the bandwidth used
between the storage backend and users, and (3) decrease latency for user
requests \cite{huang13fbcdn}. We examine the effects of our schemes
on each of these goals in turn.

\parae{Fewer Requests To The Storage Backend.} Backend storage for
images typically uses hard drives, which are capable of serving a small
number of requests per second. For instance, a typical 4TB disk holding
a large number of images is capable of a maximum of 80 Input/Output
Operations Per Second (IOPS) while keeping per-request latency
acceptably low \cite{f4}. As a result, one goal of an image caching
stack is to reduce the number of requests to the backend storage system.
Reducing requests allows images to be moved from hot storage with a high
effective replication factor to warm storage with a lower effective
replication factor sooner because the request rate would drop to what
the warm storage system could handle sooner \cite{f4}. Figure
\ref{tbl:cache_effects} shows the reduction in requests to the backend. L-ROMP, for instance, would
reduce requests to the backend by over 10\%. The reduction comes from
the hit rate increases in the caches.

\parae{Fewer Bytes To The Edge and Externally.} One of the primary goals
of Facebook's edge cache is to reduce the bandwidth required between it
and the origin cache \cite{huang13fbcdn}. ROMP reduces the
required bandwidth in two ways when deployed on the edge. First, our
recompressed images are smaller than their JPEG Optimized counterparts.
This results in a reduction in bandwidth proportional to the compression
ratio. Second, the increased cache hit ratio would lead to fewer misses
in the edge cache that need to be filled from the origin cache or the
backend. The combination of effects is shown in Figure
\ref{tbl:cache_effects}. L-ROMP, for instance, would \emph{reduce the
bandwidth between the edge and origin by 30.5\%}.

An additional benefit of using L-ROMP arises from the fact that the
image delivered to the user is often of smaller size than the uploaded
image. Thus, it can reduce the data consumption of mobile devices as
well as decrease the amount of external bandwidth of the service. This
reduction is directly proportional to the lossy savings of those
schemes, \emph{i.e.}, their compression ratio without the normal ROMP
component (Figure \ref{tbl:cache_effects}).

\begin{figure}
\centering
\noindent\includegraphics[width=.85\columnwidth]{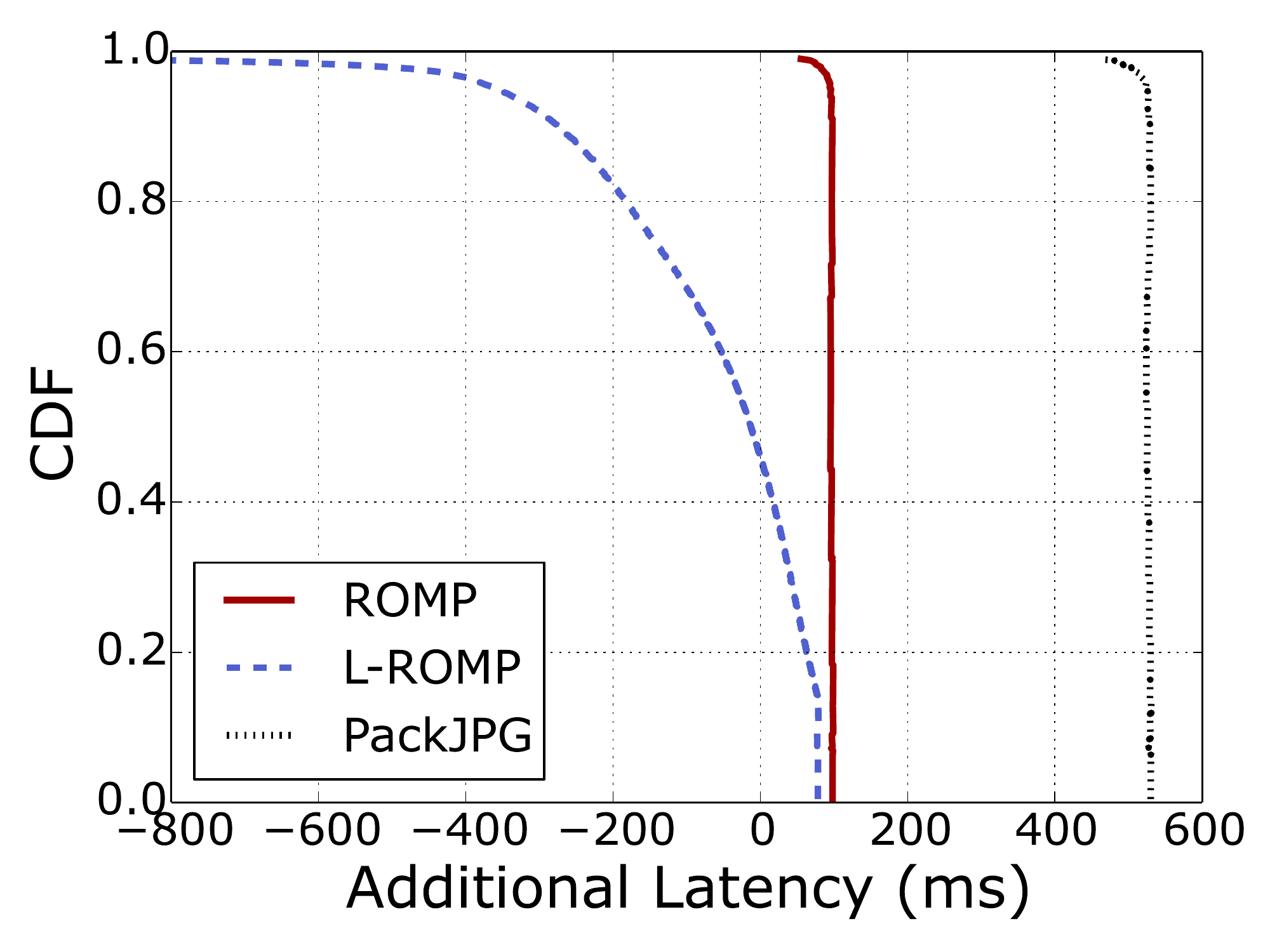}
 
\caption{Complementary cumulative distribution function (CCDF) on latency difference of deploying ROMP, L-ROMP and PackJPG.}
 
 
\label{fig:collateral_benefits_latency}
\end{figure}

\parae{Latency Effects.} One of the goals of a photo caching stack is to
decrease the latency for users to download photos. The expected latency
effects of deploying ROMP in a photo caching stack are complex and our
objective was for them to make download latency no worse, and ideally
better. The decoding time from ROMP would contribute additional latency
to every request. But, it would also reduce latency in two
ways. First, the increased hit rates at the caches would result in more
requests being served by caching layers closer to the user. Second, the
decrease in image size with L-ROMP requires fewer data transfers,
thereby reducing overall download times.

Figure \ref{fig:collateral_benefits_latency} shows the estimated latency
of downloading a $2048\times1536$ image. We see that ROMP has a
negligible effect on latency and L-ROMP reduces latency above the
40$^{\textrm{th}}$ percentile. For the tail of the distribution above
the 90$^{\textrm{th}}$ percentile, L-ROMP can reduce latency by more
than 500ms. This gain comes almost entirely from L-ROMP's reduction in
image sizes; we also estimated the latency effect for L-ROMP cache hit
rates without the image size reduction and saw a curve very similar to
ROMP. The reason the increased cache hit rates do not have a noticeable
impact is that the Facebook stack is well provisioned, \emph{i.e.},
adding extra capacity has a small effect \cite{huang13fbcdn}.

Above we mainly focused on deploying ROMP to the edge cache, the latency effect of deploying
ROMP to the transcoder tier can be implied: the additional decoding latency would only be added to the
requests served by the backend, so we observe similar latency effect
above the 70$^{\textrm{th}}$ percentile (tail 30\%),
but lower latency for other percentiles. Photo sharing services work
hard on optimizing the tail latency, and thus low
complexity codecs are required as well \cite{haystack}. Our analysis shows that
(high complexity) PackJPG inflates the latency of the fastest 40\% of
downloads by more than 50\%. Even though the decompression latency is
incurred only for cache misses, we still see significant impact on the
overall latency distribution: \tld80\% of the distribution incurs more
than 150ms additional latency.

We also estimated the collateral benefits of deploying ROMP for photo
sharing services that are not as well provisioned as Facebook,
\emph{e.g.}, services with limited cache sizes. For such services we see
slightly larger improvements in all metrics, but we omit details due to
space constraints.

%% file: related.tex
\section{Related Work}
\label{sec:related}

This section explains how our approach relates to prior work in
compressed storage systems, photo compression, and photo-sharing
architectures.

\parab{Compression in Storage Systems.}\label{compression-in-storage-systems}
The use of compression for improving the efficiency of storage systems
dates back several decades. These involve techniques for achieving
compression in file systems such as
\cite{file2,file1,file3,file4}, in databases
\cite{file5,file6} or for unstructured inputs \cite{file7}.
ROMP is inspired by this line of work, especially the idea of ``online
compression'' \cite{file1,file2,file3}, but focuses on a
relatively new class of large-scale storage systems specifically
designed for photos, whose requirements and workloads are different than
those considered by prior works. Unlike file or database compression
systems, ROMP must compress already compressed objects, leveraging the
observation that large coding tables can provide compact storage across
the entire storage system (which file and database compression systems
can leverage, but, to our knowledge, do not). In addition, our
approach requires reasoning about performance impacts across globally
distributed photo stack.

\parab{Photo Compression.}\label{photo-compression}
Compression methods for photo-sharing services need to provide high compression and low
complexity. Generic file compression techniques like gzip and bzip2,
that do not leverage specific properties of images, can only provide
negligible compression for photos beyond JPEG. The same is true of
deduplication, which has received significant attention recently
\cite{dedup1,dedup2,dedup3,dedup4,dedup5,dedup6}. We validated
this experimentally on the FiveK image set and found $\le$ 0.5\%
compression for all generic schemes we tried: gzip, bzip2, xz,
fixed-block deduplication, and variable-block deduplication.

Several papers have explored JPEG compression. Some of them focus on
compressing JPEG losslessly
\cite{lossless2,lossless1,lossless3,lossless4,packjpg-paper}.
However, as shown in Section \ref{eval}, they either cannot achieve as
high compression as ROMP, or have much higher
complexity. JPEG lossy compression methods include transcoding from JPEG
to another format (e.g., WebP \cite{webp} or JPEG2000
\cite{jpeg2000}) and transcoding to JPEG with lower quality or
resolution. The former introduces high complexity
\cite{webp,jpeg2000} and thus is not a viable option. Transcoding to JPEG but with lower quality and lower resolution
often introduces significant degradations in quality
\cite{coulombe2010transcoding}, while L-ROMP degrades more
gracefully than these approaches.

Recently, there have been several proposals for compressing photo
storage based on analyzing higher-level structures (objects, landmarks)
in similar images
\cite{cloud_compression1,cloud_compression2,cloud_compression3,cloud_compression4}.
Because they have to recognize such structural similarity, these
techniques generally have much higher complexity;
it is also not clear that their quality degradations are acceptable.

ROMP outperforms other lossless JPEG compression schemes by occupying a
unique point in tradeoff between compression and complexity with high
compression and low complexity. L-ROMP is inspired by prior work on
thresholding \cite{thresholding,thresholding2,thresholding3}, but
differs from them in only introducing perceptually lossless changes and
in its focus on low complexity as the design constraint.

\parab{Other Related Work.}\label{other-related-work}
Researchers have also explored several complementary aspects of photo
service stacks: Haystack \cite{haystack} is used for image storage
at Facebook and contains optimized metadata storage to reduce photo
fetch latency; Huang et al. \cite{huang13fbcdn} present a
measurement study of the efficacy of Facebook's distributed photo
caching architecture which resembles Figure \ref{fig:download_path}; and
f4 \cite{f4} is a storage system for photos and other binary objects
that are infrequently accessed. ROMP is complementary to this body of
work; it can be used on Haystack or f4, and can improve caches in
Facebook.

%% file: concl.tex
\section{Conclusion}
\label{sec:concl}

Motivated by the need for additional tools for managing storage in
large photo sharing services, this paper explores the problem of image
recompression in these services and proposes two low complexity
recompression schemes, ROMP and L-ROMP, that produce perceptually
lossless compression with gains of 15-28\%. Compression gains of this magnitude can substantially
reduce storage requirements at these services. In addition, they
increase cache hit ratios, reduce requests to the backend, reduce
download latency and download sizes, and reduce
wide area network traffic.